\documentstyle[preprint,aps,epsf,floats,tighten]{revtex}
\begin{document}
\draft
\title{Nonconservation of global quantum numbers in {$\overline{RS}$}-type models}
\author{Tibor Torma}
\address{Department of Physics, Oklahoma State University, 
Stillwater OK 74075}
\date{\today}
\maketitle
\begin{abstract}
In Randall-Sundrum type scenarios the effective size of the extra dimension
remains unconstrained.
A TeV-scale brane tension without orbifold boundary conditions would allow
phenomenologically observable processes at high energy colliders.
Among others, the brane could fragment into bubbles that fly away for good
or return within a time $\leq {\cal O}(1\,mm)/c$.
Particles trapped on the bubbles may fake nonconservation of global or electric
charges.

In this letter we explore the generic (model independent) features of these
bubbles. We describe the brane dynamics by a scalar field coupled derivatively
to the energy momentum tensor by dimension-8 operators.
Mass is generated to this field by any brane stabilization mechanism. 
The bubbles may be stabilized by the Casimir effect.
When the threshold energy of $\sqrt{s}\sim brane\ tension \geq TeV$ is reached,
bubbles with
$TeV^{-1}$ size are copiously produced. At lower $\sqrt{s}$, smaller bubbles
can be produced (if they exist) with strongly suppressed probability.
\end{abstract}
\narrowtext

\section*{}
Our world may be a 4-dimensional flat submanifold of a strongly curved
5-dimensional spacetime~\cite{RS1,RS2,Shape}.
One such scenario~\cite{RS1} may explain the gauge hierarchy problem when the
5-dimensional curvature radius is at the Planck scale and spacetime is a thin
``sandwich" between two 4-dimensional surfaces.
The surfaces may be stabilized at a distance somewhat exceeding the Planck
length by supposing scalar fields in the ``bulk" and proper type of
interactions~\cite{Wise}. Alternatively~\cite{RS2}, an infinite sized fifth
dimension is also conceivable with effectively 4-dimensional gravity
on one ``brane".
The bulk is a slice of $AdS_5$ with only gravity living there,
with a negative cosmological constant $\Lambda_5$.
The brane is stabilized by requiring that the metric is symmetric (``orbifold
symmetry") in the fifth coordinate.
The flatness our world is ensured by the (extreme) fine tuning of
the tension of the brane $\Lambda_4=12kM_5^3$, where $M_5$ is the 5-dimensional
Planck mass and $k^{-1}=\left(\frac{-\Lambda_5}{12M_5^3}\right)^{-1/2}$ is
the AdS length.
The resulting strength of the 4-dimensional effective gravity comes out
to be $M_4=(M_5^3/k)^{1/2}$, identified with the Planck mass $10^{19}\,GeV$.
In these models the hierarchy problem is not addressed, though.
Another version of the same set of ideas appeared in~\cite{Shape}, where it is
shown that a world confined to a third brane with small positive tension located
inside the ``sandwich" would also experience four-dimensional gravity. In that
case the ``middle" brane clearly cannot be located at an orbifold fixed point.

In these frameworks two of the three parameters ($\Lambda_5,\Lambda_4,
\mbox{and} M_5$) are fixed by the fine tuning and the observed value of
the Newtonian gravitational constant.
The third parameter is not restricted.
In the $RS_1$ framework where the aim is the hierarchy problem, it is natural to
suppose that all parameters are on the same (Planck) scale.
However, it has been speculated in~\cite{Davoud}, that the brane tension
($\Lambda_4$) could be as small as ${\cal O}(TeV)$, which corresponds to an AdS
length of $k^{-1}\sim0.1\,mm$ and such models were referred to as
$\overline{RS}$. This scale corresponds to the limit imposed by the present
experimental limits to modifications of Newtonian gravity.
A brane tension of this size is naturally expected from low-scale
supersymmetry breaking.
Coincidentally, the present limits on the observed
cosmological constant could be explained by ignoring the gravitational effect of
small-length quantum fluctuations (i.e. the coupling of their zero-point
energies to gravity), shorter than
$\sqrt{\Omega_\Lambda^{-1/2}r_{Universe}l_{Planck}}\sim{\cal O}(0.1\,mm)$. Here
$\Omega_\Lambda/r_{Universe}^2\sim+10^{-52}m^{-2}$ is the observed value of the
cosmological constant~\cite{Cosm}. We argue that such a framework is attractive
enough to consider its phenomenological consequences, even though it is somewhat
awkward to embed it in an underlying brane theory. That might require the
introduction of a very large number of coincidental branes.

In order to orient our thinking we provide one consistent set of parameters. A
five-dimensional gravity scale $M_5=5\times10^5\,TeV$ and a cosmological constant
$\Lambda_5^{1/5}=-7\,GeV$ with a brane tension $\Lambda_4^{1/4}=10\,TeV$ would
provide the correct strength four-dimensional gravity and $k^{-1}=0.1\,mm$ for
the effective size of the extra dimension.

The phenomenology of such a framework has been studied in~\cite{Davoud}. They
found no collider signature, and argued that only modifications to Newtonian
gravity on the submillimeter scale might be observable.
One might wonder however, if effects of nontrivial brane dynamics might show up
in these models. Oscillations of the brane should be suppressed only by (powers
of) the ${\cal O}(TeV)$ brane tension.
Not surprisingly we find below that such oscillations correspond to a massless
moduli field~$\Phi$ in the 4-dimensional theory.
There must be some mechanism that gives it a mass if phenomenological disaster
is to be avoided. Any mechanism of brane stabilization (such as in~\cite{Wise})
would do that.
From a phenomenological point of view, this new scalar can be
identified by the fact that it couples to the energy-momentum tensor of the
Standard Model (SM) fields through \mbox{dimension-8} operators.
The generation of topologically nontrivial $\Phi$ field configurations is also to
be considered.
In a 5-dimensional geometric language these correspond to creation of
``bubbles", whose surface is made of the same brane material as our world, in
high energy collisions through the interactions of the $\Phi$ field with the SM
particles.
Such ``bubbling" has been considered in~\cite{Dvali} in the context of the usual
$RS$ model (with Planck-scale $\Lambda_4$) as an explanation of the baryon
asymmetry of the universe.\footnote{
The instability of our world against the creation of a large number of black
holes was pointed out in~\cite{BH} when the gravity scale is lowered.
In our case, however, the brane tension is lowered to ${\cal O}(TeV)$, the bulk
gravity scale $M_5$ remains large.}

In the discussion below we will find that such bubble generation is impossible
at the $TeV$ scale in models where our brane is ``stuck" in an orbifold fixed
point of the geometry. The would-be moduli field then has mass on the Planck
scale. In the absence of such restriction however, bubbles can fly off in both
directions of the brane with equal probability.

These bubbles, once created, tend to collapse under the influence of their brane
tension. However, the vacuum energy of the fields living on the surface
increases as their size decreases (the Casimir effect) and those bubbles with
appropriate topology and periodic boundary conditions for the SM fields will be
stabilized at a size of ${\cal O}(TeV^{-1})$.
As they are pointlike on the scale of the AdS length, they will follow the
geodesics on the 5-dimensional spacetime. 
These geodesics on the far-horizon side turn back and cross our brane again
within a time and brane distance ${\cal O}(k^{-1})$.
This would lead to the observation of displaced vertices in collider experiments.
On the other hand, geodesics on the other side will accelerate away from the
brane and bubbles on this side disappear from our world.
If particles with baryon number, lepton number or electric charge are trapped on
the escaping bubble, the experiment would observe nonconservation of these
quantum numbers.
Because all this physics is on the ${\cal O}(TeV)$ scale, there is no true
violation of the quantum numbers. It has been shown in~\cite{Rubakov} that the
electromagnetic field of such disappearing charge can be consistently calculated.
We do not address possible nonconservation of color here.

The action corresponding to the model under consideration is
\begin{equation}\label{eq:1}
S=\int_{{\cal M}_5}(2M_5^3R_G-\Lambda_5)+\int_{{\cal M}_4}({\cal
L}_{SM}(\varphi,g)-\Lambda_4)
\end{equation}
where $\varphi$ represents collectively the SM fields and $g_{\mu\nu}$ is the
induced metric on the brane~${\cal M}_4$.
The metric solution to Eq.~\ref{eq:1} in the SM vacuum is $AdS_5$:
\begin{equation}
G_{ab}dx^adx^b=-dy^2+\eta_{\mu\nu}dx^\mu dx^\nu e^{-2ky}.
\end{equation}
The equation for the geodesics is 
\begin{equation}
x^\mu(y)={1\over k}n^\mu \sqrt{e^{2ky}-e^{2ky_0}}
\end{equation}
where $n^\mu$ is a timelike unit vector ($n^\mu n^\nu\eta_{\mu\nu}=1$). The
geodesic (see Fig.~1) has a turning point at some $y_0<0$. The time it takes for
the geodesics to return to the brane is readily calculated
\begin{equation}
t={1\over k}\,\frac{2\gamma\sin\alpha}{\sqrt{1+\gamma^2\sin^2\alpha}}
\end{equation}
where $\alpha$ is the angle of the initial speed to the brane and $1-1/\gamma^2$
is the initial 5-speed of the geodesic.
As in an accelerator experiment we do not expect extremely ultrarelativistic
bubbles, this time is in the submillimeter range.

\begin{figure}[htb]
\begin{center}
\fbox{
\epsfxsize=9cm\epsfbox{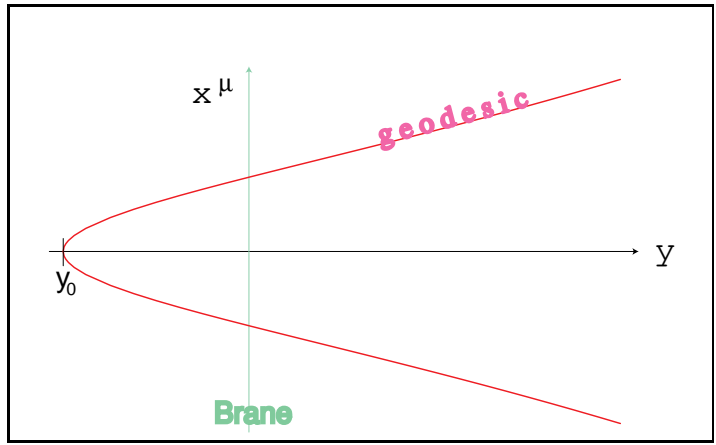}
}
\end{center}
\caption{\protect\label{fig:1} The bubbles once generated follow the  geodesics
of $AdS_5$. The AdS horizon is to the right.}
\end{figure}

The classical equations of motion of the brane can be readily derived, at least
formally, from Eq.~\ref{eq:1}.
Choosing coordinates $x^\mu$ on the brane and
varying the position $X^a(x)$ of the brane, as well as the metric $G_{ab}$, in
the action we find
\begin{eqnarray}
\left(
T^{\mu\nu}-\Lambda_4\,g^{\mu\nu}
\right)\left(
D_\mu D_\nu X^c+\Gamma^c_{ab}\partial_\mu X^a\partial_\nu X^b
\right)&=&0\\
\frac{1}{2\sqrt{-G}}\int
d^4x\sqrt{-g(x)}\,\delta^5\left[z-X(x)\right]\,
\left(T^{\mu\nu}-\Lambda_4g^{\mu\nu}\right)
\partial_\mu X^a\partial_\nu X^b
&=&
2M_5^3E^{ab}+{1\over2}G^{ab},
\nonumber
\end{eqnarray}
where the first equation is satisfied on the brane (${\cal M}_4$) and the
second everywhere on (${\cal M}_5$). The quantity $T^{\mu\nu}$ is the Standard
Model energy-momentum tensor; $E^{ab}$ is the Einstein tensor calculated from
the metric $G^{ab}$.

The first equation is a variant of the geodesic equation. The fact that it is
invariant under a rescaling of all energy-momentum on the brane
($T^{\mu\nu}\rightarrow\lambda T^{\mu\nu}$,
$\Lambda_4\rightarrow\lambda\Lambda_4$) is a reflection of the equivalence
principle.

In Gaussian normal coordinates of the brane (i.e. $X^y(x)=0$ and
$X^\alpha(x)=x^\alpha$) the second equation becomes
\begin{equation}\label{eq:6}
2M_5^3\,E^{ab}+{1\over2}\Lambda_5G^{ab}=-{1\over2}\delta(y)
\left(T^{\mu\nu}-\Lambda_4g^{\mu\nu}\right).
\end{equation}
This can be replaced by the vacuum equations for $y\neq0$ and the Israel
conditions~\cite{Israel}
\begin{equation}\label{eq:isr}
\left(\lim_{y\to+0}-\lim_{y\to-0}\right)\,\left(K_{\mu\nu}-K\,g_{\mu\nu}\right)
=-\frac{1}{4M_5^3}
\left(T^{\mu\nu}-\Lambda_4g^{\mu\nu}\right).
\end{equation}
Here $K_{\mu\nu}$ is the extrinsic curvature of the brane in the metric
$G_{ab}$.
The first equation becomes
\begin{equation}\label{eq:rep}
\left(T^{\mu\nu}-\Lambda_4\,g^{\mu\nu}\right)
K_{\mu\nu}=0.
\end{equation}
This equation is ambiguous when no orbifold symmetry is imposed
($y\leftrightarrow-y$). A mathematically careful treatment of the variational
problem in Eq.~1 reveals that the correct equation is found if $K_{\mu\nu}$ is
replaced by the average of the two $K_{\mu\nu}$'s on the twe sides of the brane
in Eq.~\ref{eq:rep}.

In collider experiments we probe the regime where $T^{\mu\nu}$ is much smaller
than $M_5$. We can see in Eqns.~\ref{eq:6},\ref{eq:isr},\ref{eq:rep} that the
$\frac{1}{M_5}\rightarrow0$ limit must be taken. Then the metric $G^{ab}$ does
not feel the effect of the brane any more and is exactly $AdS_5$, with the brane
moving on this background, subject to only Eq.~\ref{eq:rep}.

For a brane subject to orbifold boundary conditions, such fixed background
prohibits all local perturbations. (This situation is similar to what happens on
a flat background. Then the fixed surfaces of $Z_2$ isometries are only flat
planes.) All processes involving bubble production (or production of the moduli
for that matter) would be subleading in $\frac{energy}{M_5^3}$, suppressed by 15
orders of magnitude in our numerical example.

When no symmetry requirement is imposed, it is reasonable to write the equations
of motion in flat 5-dimensional coordinates and specify the brane by one
function (normalized so for future convenience):
\begin{equation}
X^\mu(x)=x^\mu\ \ \ \mbox{and}\ \ \ X^5=\Lambda_4^{-{1\over2}}\Phi(x).
\end{equation}
The induced metric in these coordinates becomes
\begin{equation}
g_{\mu\nu}=\eta_{\mu\nu}-\frac{1}{\Lambda_4}\partial_\mu\Phi\partial_\nu\Phi,
\ \ \ \mbox{and also}\ \ \ 
\sqrt{-g}=\sqrt{1-\frac{1}{\Lambda_4}
\eta^{\mu\nu}\partial_\mu\Phi\partial_\nu\Phi}.
\end{equation}
The bulk part of the action is not varied any more but the brane part becomes
\begin{equation}\label{eq:action}
S=\int d^4x\,{\cal L}\ \ \ \mbox{with}\ \ \ {\cal L}(\Phi,\varphi)=
\sqrt{-g}\,\left[{\cal L}_{SM}(g_{\mu\nu},\varphi)-\Lambda_4\right].
\end{equation}
Expanding this Lagrangian in the $\Phi$ field we see that is is canonically
normalized and is coupled derivatively to the energy momentum tensor of the SM
fields. It is not hard to see that all the couplings, both to fermions and
bosons, are dimension 8 and higher. For example, to a scalar we have a coupling
$-{1\over2\Lambda_4}\left(\partial_\mu\varphi\,\partial^\alpha\Phi\right)^2$.
The symmetry $\Phi\rightarrow\Phi+const.$ corresponds to the translational
invariance of the brane in the fifth direction. The $\Phi$ field is then the
Goldstone boson corresponding to this symmetry. 

Such a massless moduli field with $TeV$-scale couplings phenomenologically
unacceptable. However, its Goldstone nature also implies that any explicit
breaking of the fifth dimensional translations would give mass to the $\Phi$
field. In particular, so do mechanisms that stabilize interbrane separations in
models with more than one brane~\cite{Wise}. In this letter we are only trying to
establish the generic consequences of the $\overline{RS}$ idea and simply
suppose that one way or another such mass has been generated.

Bubbles of brane matter can be considered as nontrivial topological
configurations of the $\Phi$ fields. The classical equations of motion 
can be found in terms of the $\Phi$ field from the action in Eq.~\ref{eq:action}.
Neglecting the SM fields,
\begin{equation}
\partial_\mu\frac{\eta^{\mu\nu}\partial_\nu\Phi}{
\sqrt{1-{1\over\Lambda_4}\eta^{\alpha\beta}
\partial_\alpha\Phi\partial_\beta\Phi}}
=0.
\end{equation}
For small field strength these equations describe harmonic oscillations of the
brane.

\begin{figure}[htb]
\begin{center}
\fbox{
\epsfysize=3cm\epsfbox{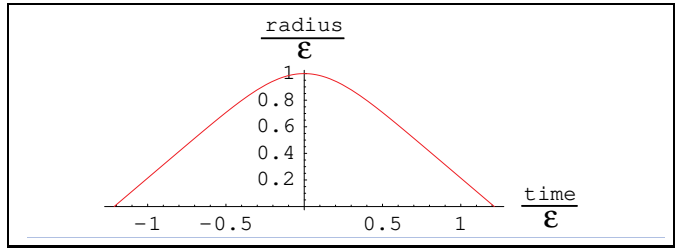}
}
\end{center}
\caption{\protect\label{fig:2} The classical collapse of a bubble. Here,
${\cal E}^3=\frac{2}{\pi^2}\frac{Energy}{\,\Lambda_4}$ determines the lifetime.}
\end{figure}

Classically, these equations can be solved supposing spherical symmetry. The
result is a collapsing bubble with time dependent radius $R(t)$, shown in
Fig.~\ref{fig:2},
\begin{equation}
t(R)=t_0\pm{\cal E}\int_0^{R/\cal E}\frac{du}{\sqrt{1-u^6}}.
\end{equation}
The lifetime of the bubble equals $2.43\,{\cal E}$, where ${\cal
E}^3=\frac{2E}{\pi^2\Lambda_4}$, with $E$ being the total energy contained in
the bubble.
This result is physically not surprising.
When the SM field are thrown
away, there is nothing to counteract the brane tension in pulling together the
bubble.

Including the quantum fields on the surface of the bubble, the Casimir effect
generates a vacuum energy density $\sim-\frac{\alpha}{R^4}$. The value and sign
of the constant $\alpha\leq{\cal O}(1)$ depends on the topology of the bubble
and the boundary conditions (periodic vs. antiperiodic), and is generically
unknown. In a supersymmetric theory it may cancel~\cite{Milton} but even in the
simple case of QED on $R\times S^3$ its sign is presently
debated~\cite{Ravndal}. 
A positive value of $\alpha$ results in a stabilized bubble within a very short
time $\sim{\cal E}$, with the Casimir energy roughly equal to the energy
contained in the brane surface:
\begin{equation}
\frac{\alpha}{R^4}\times \frac{\pi^2}{2}R^3\approx\frac{\pi^2}{2}\Lambda_4R^3,
\ \ \ \mbox{i.e.}\ \ \ \Lambda_4R^4\approx\alpha
\end{equation}
This fixes the internal energy contained in the bubble at the $TeV$ scale.
We may take the point of view that bubbles with all sorts of topologies will be
created but only those with a positive $\alpha$ survive.

The situation is analogous to perturbative bosonic string theory. In that case
the classical string also collapses. In quantum language this is due to the
presence of tachionic excitations. With more fields present on the worldsheet
it is possible to arrange a situation when the tachionic modes decouple and the
result is a spectrum with energy levels$\sim \Lambda_2^{1/2}$. In our case it is
impossible to show without a detailed discussion of the field theory on the brane
how such a mechanism explicitly works.

Now we turn to the discussion of bubble production.
It should also be noted first that the quanta of the $\Phi$ field can be produced
in pairs, due to the $\Phi\rightarrow-\Phi$ symmetry.
This is a remnant of the $y\rightarrow-y$ symmetry of the Lagrangian which
remains even if the metric solution is not required be symmetric.
As a consequence, the probability to produce a bubble will be equal on the two
sides. The difference shows up only at the scale of the $AdS$ curvature, well
separated from the scale of bubble production.

In classical physics it is impossible to find the probability of producing a
bubble in the linear approximation. The {\it in} state is vacuum and the
$\Phi\leftrightarrow-\Phi$ symmetry prohibits any solution with $\Phi\neq0$. Any
solution arises as a consequence of the instability of the $\Phi=0$ solution.
Even if we find such an unstable solution, the response is proportional to the
input uncertainty of the $\Phi$ field strength (i.e. the brane position) which
is not described correctly by classical physics.

\begin{figure}[htb]
\begin{center}
\fbox{
\epsfysize=6cm\epsfbox{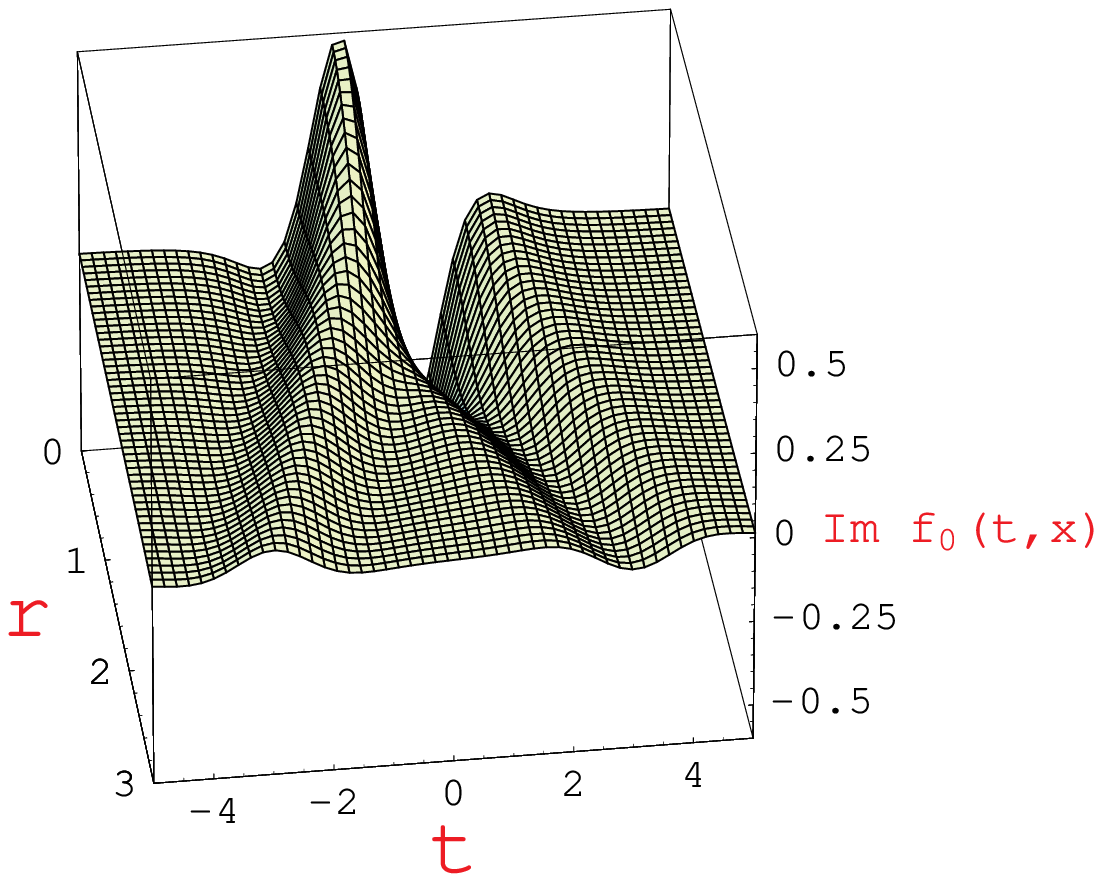}}
\fbox{
\epsfysize=6cm\epsfbox{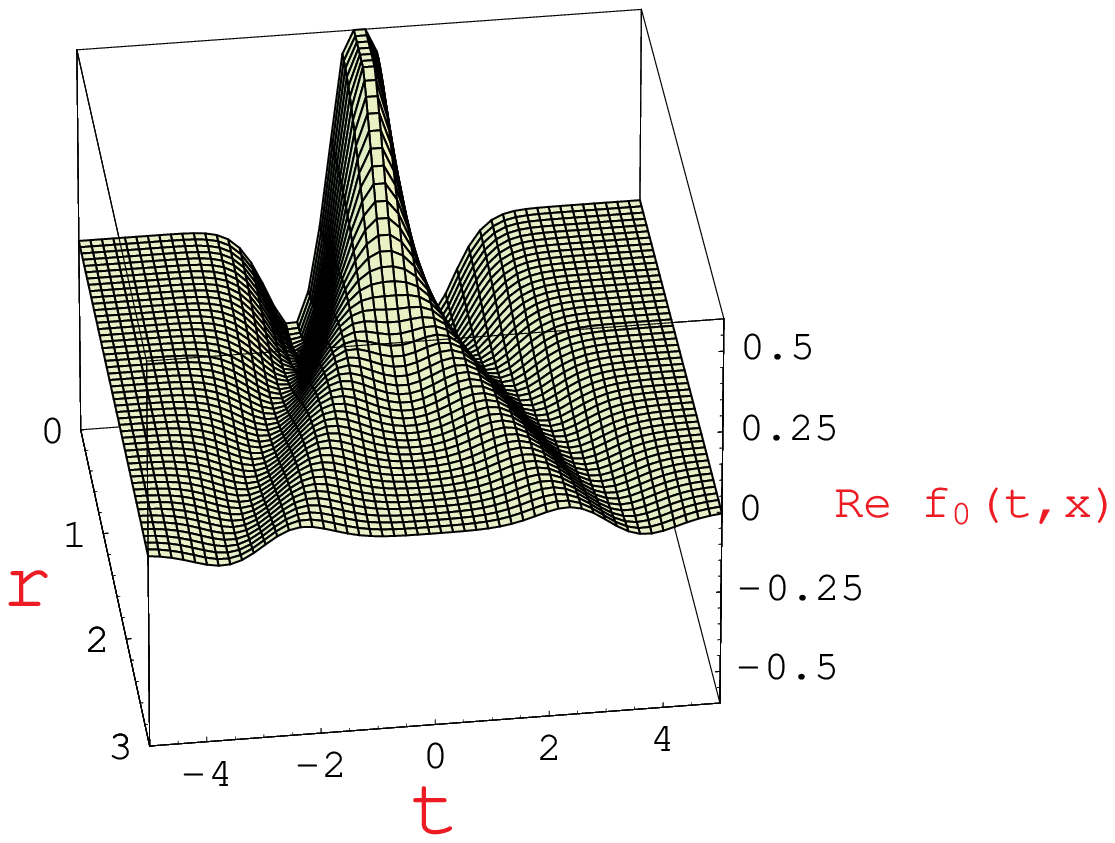}
}
\end{center}
\caption{\protect\label{fig:3} The time evolution of the wave function of the
particle modeling the SM process. The m=0 case is shown.}
\end{figure}

In order to provide at least a crude quantum description of bubble generation we
first model the SM process.
For technical simplification we look at (3-dimensional) spherical symmetry.
We take a state that contains one free particle localized in a region with
radius~$\rho$ that is allowed to flow apart in time to all directions. The wave
function at $t=0$ is chosen to be
$f(t=0,{\bf x})=\left(\frac{2}{\pi\rho^2}\right)^{3/4}e^{-\frac{r^2}{\rho^2}}$.
The time evolution of this state is trivially found from the Fourier transform
of the wave function which satisfies the Klein-Gordon equation (See
Fig.~\ref{fig:3}.)

The energy-momentum tensor is calculated next
\begin{equation}\label{eq:14}
T_{\mu\nu}=\left(\frac{\rho^2}{2\pi}\right)^{3/2}
\int\int\frac{d^3{\bf k}\,d^3{\bf q}}{(2\pi)^3\sqrt{E_{\bf k}E_{\bf q}}}
\left[
k_{(\mu}q_{\nu)}+\eta_{\mu\nu}\frac{(k-q)^2}{4}
\right]
e^{
-{\rho^2\over4}\left({\bf k}^2+{\bf q}^2\right)
}
\cos\left[
t(E_{\bf k}-E_{\bf q})-{\bf x}({\bf k}-{\bf q})
 \right].
\end{equation}

In the next step we calculate the two particle state that is produced in the
lowest order in ${1\over\Lambda_4}$ in perturbation theory. Note that although
perturbation theory is not expected to be a good guide in the strong field
regime, its predictions will provide the correct order of magnitude. The state
at time $t$ is $|t\!>\ =|0\!>+\ |t\!>_2\ $ with
$|t\!>_2=\,
\int\frac{d^3{\bf k}_1}{(2\pi)^{3/2}\sqrt{2E_{{\bf k}_1}}}
\int\frac{d^3{\bf k}_2}{(2\pi)^{3/2}\sqrt{2E_{{\bf k}_2}}}
c({\bf k}_1,{\bf k}_2)\hat{a}^\dagger_{{\bf k}_1}\hat{a}^\dagger_{{\bf k}_1}
|0\!>
$ and
\begin{equation}\label{eq:15}
c({\bf k}_1,{\bf k}_2)\,=\frac{-i}{2\Lambda_4}
\int_{-\infty}^tdt^\prime\int d^3{\bf x}^\prime
e^{i{\bf x}^\prime({\bf k}_1+{\bf k}_2)}
\left[
T_{00}(t^\prime,{\bf x}^\prime)E_{{\bf k}_1}E_{{\bf k}_2}-T_{jl}(t^\prime,{\bf
x}^\prime)k_1^jk_2^l
\right].
\end{equation}
where the free-field operator generates an elementary excitation of the $\Phi$
field.

The actual calculation of the probability to produce a topologically nontrivial
field configuration requires exact knowledge of the fields on the brane.
This is an unsolved problem beyond the scope of the present paper.
Instead, we conjecture that a bubble of size $R$ will form whenever the energy
density is larger than $\Lambda_4$ in a region of size $R$. If that happens,
than with ${\cal O}(1)$ probability we will have a fluctuation of that
compensates the force that is pulling together the brane. The intuitive
classical image is that when the surface tension of the brane is compensated in
a region, an instability develops and the brane will bulge out. Such a would-be
bubble, if long enough compared to its size, will rather detach than be pulled
back into our world.

Next we calculate the average energy density carried by the $\Phi$ field. We find
\begin{eqnarray}\label{eq:16}
<t|{\cal H}(t,{\bf x})|t>&=&
\int\frac{d^3{\bf q}_1}{(2\pi)^32E_{\bf{q}_1}}  
\int\frac{d^3{\bf q}_2}{(2\pi)^32E_{\bf{q}_2}}  
\left(E_{{\bf q}_1}q_2^j+E_{{\bf q}_2}q_1^j\right)
\\&&\times
\left[c({\bf q}_1,{\bf q}_2)e^{i\,x\cdot(q_1+q_2)}+
 \overline{c({\bf q}_1,{\bf q}_2)}e^{-i\,x\cdot(q_1+q_2)}\right]
+{\cal O}\left({1\over\Lambda_4^2}\right).
\nonumber
\end{eqnarray}

When we substitute Eqns.~\ref{eq:14},\ref{eq:15} into Eq.~\ref{eq:16}, we find a
short-distance divergence. We regulate it by calculating the total amount of
energy contained within radius $R$,
\begin{eqnarray}
{\cal E}(R)&\equiv&\int d^3{\bf x}<t|{\cal H}(t,{\bf x})|t>\\
&=&\frac{(\rho R)^3}{4\pi\Lambda_4}\int_0^\infty dp p^4
e^{-\frac{\rho^2+R^2}{8}p^2}\int_{-\infty}^\infty dE e^{-itE}
{\tilde T}^E(p)\ln(p-E+i0)+c.c.\nonumber
\end{eqnarray}
with the $\Phi$ mass neglected, $E_\pm=|{\bf k}\pm\frac{1}{2}{\bf p}|$, and
\begin{eqnarray}
{\tilde T}^E(p)&=&{1\over3}\int\frac{d^3{\bf k}}{\sqrt{E_+E_-}}
e^{-\frac{\rho^2}{2}{\bf k}^2}\delta(E_+-E_-E)
\left(
(E_+-E_-)^2\ \left\{8{\bf k}^2-7\frac{(\bf k\cdot p)^2}{{\bf p}^2}-
  \frac{11}{4}(E_+-E_-)^2+{\bf p}^2   \right\}\right.\nonumber
\\&&\left.
-{\bf p}^2\ \left\{4{\bf k}^2-8\frac{(\bf k\cdot p)^2}{{\bf p}^2}-
  \frac{13}{4}(E_+-E_-)^2+2{\bf p}^2   \right\}
\right).
\end{eqnarray}

\begin{figure}[htb]
\begin{center}
\fbox{
\epsfxsize=9cm\epsfbox{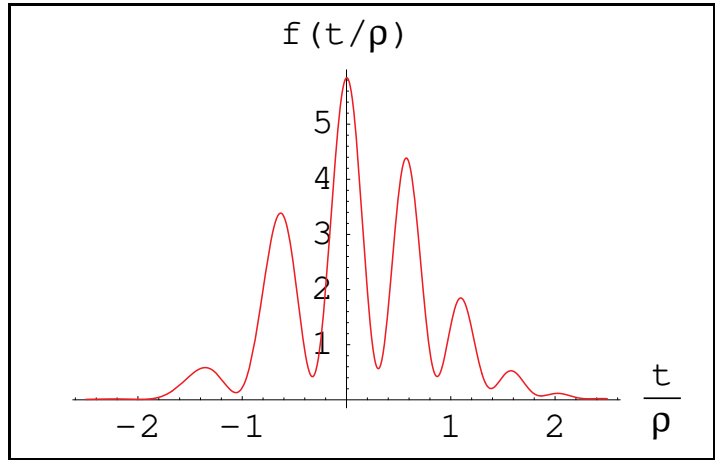}
}
\end{center}
\caption{\protect\label{fig:4} The energy density in the $\Phi$ field at $R=0$ as
a function of time.
The normalization is according to
$\rho^8\Lambda_4\frac{{\cal E}(R)}{R^3}=f\left(\frac{t}{\rho}\right)$.}
\end{figure}

The resulting averaged energy density is shown in Fig.~\ref{fig:4} in the case
when $R<<\rho$.
The production probability of a bubble is ${\cal O}(1)$ during time $R$ in a
region of size $R$ wherever this energy density is in excess of $\Lambda_4$.
Bubbles with size $R>>\rho$ can never be generated (With
$\rho^4\leq\Lambda_4^{-1}$, the function $f$ is never $>>1$, and for large $R$ it
is more suppressed.) The shape of the function in Fig.~\ref{fig:4} tells us that 
to produce a bubble at all we need a minimum energy concentration
$\rho^4\Lambda_4\leq\sqrt{f(0)}$. Once we achieve that, bubbles of all sizes
$R\leq\rho$ will be produced with unit probability (this is a sign of that in
the ``hot" region of size $\rho$ an instability occurs.)

We see now that copious bubble production requires that the incoming particles be
localized in a region $\rho\sim TeV^{-1}$.
In a particle beam $\rho$ corresponds to the beam energy,
$\rho\sim{1\over\sqrt s}$.
The above calculation shows is that the threshold of bubble production is at
$\sqrt{s}\sim\Lambda_4^{-1/4}$.
This is so even for bubbles {\it smaller} than $R<\Lambda^{-1/4}$ whose
production is possible only through tunneling.
Even though the energy might be available in processes with
$\sqrt{s}\sim\Lambda_4R^3<\Lambda_4^{-1/4}$, an energy density $\Lambda_4$ is
not attained. Once the above threshold is reached, the production
probability is ${\cal O}(1)$, the cross section is determined by the unitarity
bound $\sigma\approx{1\over s}$.

The above rudimentary calculation of the {\it expectation value} of the energy
density disregards energy fluctuations. Below the above threshold, with a small
probability, it is possible that a large energy density fluctuation arises in a
small region $R^3$ whose lifetime will be ${\cal O}(R)$.
The probability of this is the same as producing a state of mass $M\sim1/R$
(so it is localized in the bubble) and energy $E\sim{\Lambda_4 R^3}$.
This state represents the ``hot region" mentioned in the previous discussion,
and will decay into a bubble with branching ratio
${\cal O}(\frac{E^4}{\Lambda_4})\sim(\Lambda_4R^4)^3$), due to the derivative
couplings of the $\Phi$ field.
This state is far off-shell and this introduces a suppression
$(\frac{E}{{M}})^4\sim(\Lambda_4R^4)^4$, so the cross section is
$\sigma\sim\frac{(\Lambda_4R^4)^7}{s}$. If we look at the largest bubbles that
can be produced (they have the largest probability),
$\sqrt{s}\sim E\sim\Lambda_4R^3$, we find
$\sigma\sim\frac{1}{s}\,\left(\frac{\sqrt{s}}{\Lambda_4^{1/4}}\right)^{28/3}$.

We thus find that a high power of the energy suppresses the production of small
bubbles. These bubbles in fact correspond to the massless states in string theory
but in our case their mass (or existence) can be established only in a detailed
model. If they exist, they should be visible in accelerator experiments somewhat
below $\sqrt{s}<\Lambda_4^{1/4}.$ They would show up most easily if a charged
particle gets trapped on the bubble.

Note that this effect does not induce (apparent) proton decay. Even though
bubbles may be produced with a tiny probability at energies less than a $GeV$,
no baryons can be trapped on them due to energy conservation. Ultimately, this
is due to the fact that the bubble is not vanishing, only stopping to interact
with our world.

Note that in the cosmological context at present time, even in supernova
explosions the available energy for bubble production is too small. With a
reasonable choice of $\sqrt{s}\sim20MeV$ during the explosion, the average time a
proton in a neutron star needs to form a bubble is (according to the above cross
section) $\sim 10^{9}$ years, while the explosion lasts only tens of seconds.

In sum, the expected experimental signature includes (i) displaced vertices due
to returning bubbles, (ii) missing energy and/or $p_T$, as well as missing
baryon number, lepton number and electric charge as soon as the total energy
reaches $\sqrt{s}\sim\Lambda_4^{1/4}$.

\section*{Acknowledgments}

The author is grateful to Drs. S. Nandi and K. Babu for numerous discussions.

\end{document}